\documentclass[doublecol]{epl2}

\usepackage{amsmath}
\usepackage{amsfonts}
\usepackage{color}
\usepackage[colorlinks,bookmarks=false,citecolor=blue,linkcolor=red,urlcolor=blue]{hyperref}
\usepackage{ulem}

\usepackage{graphicx,subfigure}

\newcommand{\red}[1]{\textcolor[rgb]{1.00,0.00,0.00}{#1}}

\def \be {\begin{equation}}
\def \ee {\end{equation}}
\def \ba {\begin{array}}
\def \ea {\end{array}}
\def \bea {\begin{eqnarray}}
\def \eea {\end{eqnarray}}

\def \ble {\begin{widetext}\begin{equation}}
\def \ele {\end{equation}\end{widetext}}
\def \blea {\begin{widetext}\begin{eqnarray}}
\def \elea {\end{eqnarray}\end{widetext}}

\def \nn {\nonumber}

\def \a {\alpha}

\def \g {\gamma}

\def \d {\delta}
\def \D {\Delta}

\def \ve {\varepsilon}

\def \lam {\lambda}
\def \s {\sigma}

\def \r {\rho}

\def \f {\frac}

\def \lag {\langle}
\def \rag {\rangle}

\def \tr {\mathrm{tr}}

\def \cF {{\mathcal F}}

\def \univ {{\textrm{univ}}}

\def \sr {\sqrt}
\def \ii {{\mathrm{i}}}

\title{Universal R\'enyi entanglement entropy of quasiparticle excitations}
\author{Jiaju Zhang\inst{1,2} \and M.~A.~Rajabpour\inst{3}}
\shortauthor{J.~Zhang and M.~A.~Rajabpour}

\institute{
  \inst{1} Center for Joint Quantum Studies and Department of Physics, School of Science, Tianjin University, 135 Yaguan Road, Tianjin 300350, China\\
  \inst{2} SISSA and INFN, Via Bonomea 265, 34136 Trieste, Italy\\
  \inst{3} Instituto de Fisica, Universidade Federal Fluminense, Av.~Gal.~Milton Tavares de Souza s/n, Gragoat\'a, 24210-346, Niter\'oi, RJ, Brazil
}

\abstract{
  The R\'enyi entanglement entropies of quasiparticle excitations in the many-body gapped systems show a remarkable universal picture which is independent of the model, the quasiparticle momenta, and the connectedness of the subsystem. In this letter we calculate exactly the single-interval and double-interval R\'enyi entanglement entropies of quasiparticle excitations in the many-body gapped fermions, bosons, and XY chains and find additional contributions to the universal R\'enyi entanglement entropy in the excited states with quasiparticles of different momenta. The additional terms are different in different models, depend on the momentum differences of the quasiparticles, and are different for the single interval and the double interval. We derive the analytical R\'enyi entanglement entropy in the extremely gapped limit, matching perfectly the numerical results as long as either the intrinsic correlation length of the model or all the de Broglie wavelengths of the quasiparticles are small. When the momentum difference of any pair of distinct quasiparticles is small, the additional terms are non-negligible. We argue that the derived formulas can be applied for a wider range of models than those discussed here.
}

\begin{document}

\maketitle

%\end{CJK*}

%\tableofcontents

%\paragraph{Introduction.}

%\tableofcontents

%\section{Introduction}

Using entanglement in the study of quantum many-body systems has been very fruitful in the last couple of decades. In particular, different measures of entanglement have been used extensively to study quantum phase transitions and different phases of the matter \cite{Amico:2007ag,Eisert:2008ur,calabrese2009entanglement,Laflorencie:2015eck}. Not surprisingly, most of these studies were about the entanglement content of the ground state of the quantum many-body systems. There are also many studies regarding the entanglement in the excited states which one can divide to at least two categories.
The first group studies the entanglement entropy in the highly excited states which is relevant in the study of local thermalisation after quantum quenches, see \cite{DAlessio:2016rwt} and references therein. It is expected that the entanglement entropy in this regime typically follows the volume-law \cite{Deutsch:2013Microscopic,Santos:2012Weak,Beugeling:2015Global,Garrison:2015lva}, see also \cite{Vidmar:2017uux,Vidmar:2017pak,Huang:2019dxk,Vidmar:2018rqk,Lydzba:2020qfx} for the behavior of the average entanglement entropy of the excited states. The second group studies the low-lying excited states which is more relevant for the investigation of low temperature behavior \cite{Alba:2009th,Alcaraz:2011tn,Berganza:2011mh,Pizorn2012Universality,Essler2013ShellFilling,Berkovits2013Twoparticle,Storms2014Entanglement,%
Calabrese2014Entanglement,Molter2014Bound,Taddia:2016dbm,Castro-Alvaredo:2018dja,Castro-Alvaredo:2018bij,Castro-Alvaredo:2019irt,Castro-Alvaredo:2019lmj,%
Jafarizadeh:2019xxc,You:2020osa}.
In this regime one is interested in the excess entropy of the low-lying states compared to that of the ground state.
When the system is integrable or in general when there is no particle production one can consider quasiparticle excitations whose entanglement shows a remarkable universal property which has a natural qubit interpretation \cite{Castro-Alvaredo:2018dja,Castro-Alvaredo:2018bij}, see also \cite{Pizorn2012Universality,Berkovits2013Twoparticle,Molter2014Bound}.
In \cite{Castro-Alvaredo:2018dja,Castro-Alvaredo:2018bij,Castro-Alvaredo:2019irt} it was shown that the entanglement entropy for the excited  states  composed  of  finite  numbers  of quasiparticles with finite De Broglie wavelengths or finite intrinsic correlation length is largely independent  of  the  momenta  and  masses  of  the  excitations,  and  of  the  geometry,  dimension  and connectedness  of  the  entanglement  region.

In this letter we calculate exactly the single-interval and double-interval R\'enyi entanglement entropies of quasiparticle excitations in the  many-body gapped fermions, bosons, and XY chains beyond the regimes that were considered in \cite{Castro-Alvaredo:2018dja,Castro-Alvaredo:2018bij,Castro-Alvaredo:2019irt}.
For the excited states with more than one mode excited we prove that in the most generic circumstances apart from the universal terms there are extra terms.
Especially, these additional terms have nontrivial dependence on the momenta of the excited quasiparticles.
These terms are non-negligible when the momentum difference of at least a pair of distinct quasiparticles is small. Moreover, we argue that these extra terms are also universal though in a weaker sense.
In the case of double intervals in the XY chain we find a novel universal formula.
In all the cases we confirm the validity of our exact results by using exact numerical calculations.
The letter is organized as follows: we first list the essential definitions and the setup of our work. We then briefly review the universal entanglement discovered in \cite{Castro-Alvaredo:2018dja,Castro-Alvaredo:2018bij,Castro-Alvaredo:2019irt}. Then we show our results regarding the single-interval and double-interval R\'enyi entanglement entropies of quasiparticle excitations in the many-body gapped fermions, bosons, and XY chains. Finally we comment on the universality and possible applications of our results.

\begin{figure}[ht]
  \centering
  \includegraphics[width=0.375\textwidth]{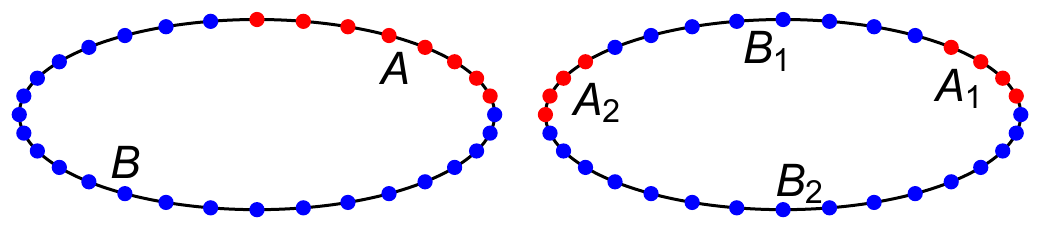}\\
  \caption{The subsystem of a single interval (left) and the subsystem of double interval (right) on a circular chain of $L$ sites. For the single interval $A$ the length is $|A|=\ell$, and for the double interval $A=A_1\cup A_2$ the lengths are $|A_1|=\ell_1$, $|A_2|=\ell_2$, $|B_1|=d_1$, $|B_2|=d_2$. For the single interval one can define a useful parameter $x=\f{\ell}{L}$, and for the double interval the parameters $x_1=\f{\ell_1}{L}$, $x_2=\f{\ell_2}{L}$, $y_1=\f{d_1}{L}$, $y_2=\f{d_2}{L}$, $x=x_1+x_2$, $y=y_1+y_2$.}\label{subsystems}
\end{figure}

Consider a quantum system in a pure state $\r_\psi=|\psi\rag\lag\psi|$, then divide it into a subsystem $A$ and its complement $B$.
Tracing out the degrees of freedom of $B$, one gets the reduced density matrix (RDM) of the subsystem $A$, i.e. $\r_{A,\psi}=\tr_B \r_\psi$.
Then the R\'enyi entanglement entropy is defined as
\be
S_{A,\psi}^{(n)} = -\f{1}{n-1} \log \tr_A\r_{A,\psi}^n.
\ee
The von Neumann entanglement entropy (the von Neumann entropy) $S_{A,\psi} = -\tr_A(\r_{A,\psi}\log\r_{A,\psi})$ can be derived by taking the $n\to1$ limit of the R\'enyi entanglement entropy.
In this letter we study the entanglement entropy of a single interval and two disjoint intervals on a circular chain of length $L$ as depicted in FIG.~\ref{subsystems}.
%For the case of the single interval $A$ there is length $|A|=\ell$, and for the double interval $A=A_1\cup A_2$ there are $|A_1|=\ell_1$, $|A_2|=\ell_2$, $\ell=\ell_1+\ell_2$.
The single-interval von Neumann and R\'enyi entanglement entropies in various many-body systems were calculated in the ground state \cite{Bombelli:1986rw,Srednicki:1993im,chung2001density,Vidal:2002rm,peschel2003calculation,Latorre:2003kg,Plenio:2004he,Cramer:2005mx}
and excited states \cite{Alba:2009th,Alcaraz:2011tn,Berganza:2011mh,Pizorn2012Universality,Essler2013ShellFilling,Berkovits2013Twoparticle,Storms2014Entanglement,%
Calabrese2014Entanglement,Molter2014Bound,Taddia:2016dbm,Castro-Alvaredo:2018dja,Castro-Alvaredo:2018bij,Castro-Alvaredo:2019irt,Castro-Alvaredo:2019lmj}.
The double-interval von Neumann and R\'enyi entanglement entropies were studied in
\cite{Furukawa:2008uk,Facchi:2008Entanglement,Caraglio:2008pk,Alba:2009ek,Igloi:2009On,Fagotti:2010yr,Alba:2011fu,%
Rajabpour:2011pt,Coser:2013qda,DeNobili:2015dla,Coser:2015dvp,Ruggiero:2018hyl}.
Following \cite{Alcaraz:2011tn,Berganza:2011mh}, we use $\cF_{A,\psi}^{(n)}$ to denote the difference between the R\'enyi entanglement entropy $S_{A,\psi}^{(n)}$ in the excited state $|\psi\rag$ and the R\'enyi entanglement entropy $S_{A,G}^{(n)}$ in the ground state $|G\rag$ as
\be
S_{A,\psi}^{(n)} = S_{A,G}^{(n)} - \f{1}{n-1} \log \cF_{A,\psi}^{(n)},
\ee
where
\be
\cF_{A,\psi}^{(n)} = \f{\tr_A\r_{A,\psi}^n}{\tr_A\r_{A,G}^n}.
\ee
{We call $\cF_{A,\psi}^{(n)}$ the R\'enyi entanglement entropy power.}%

The universal von Neumann and R\'enyi entanglement entropies in states of quasiparticle excitations were investigated in \cite{Castro-Alvaredo:2018dja,Castro-Alvaredo:2018bij,Castro-Alvaredo:2019irt,Castro-Alvaredo:2019lmj}, see also \cite{Pizorn2012Universality,Berkovits2013Twoparticle,Molter2014Bound}.
One can denote the general quasiparticle excited state as $|k_1^{r_1}k_2^{r_2}\cdots k_s^{r_s}\rag$.
There are $s$ different quasiparticles with momenta $k_i$, $i=1,2,\cdots,s$, and for each kind of quasiparticle with momentum $k_i$ there are $r_i$ excitations.
There are totally $R=\sum_{i=1}^s r_i$ excited quasiparticles.
%We use the convention that the momenta $k_i$ are integers or half-integers depending on the boundary conditions of the circular chain, and the physical momenta of the quasiparticles are $p_i=\f{2\pi k_i}{L}$.
{We use the convention that the momenta $k_i$ are integers or half-integers depending on the boundary conditions of the circular chain, which is more convenient than convention in \cite{Castro-Alvaredo:2018dja,Castro-Alvaredo:2018bij,Castro-Alvaredo:2019irt,Castro-Alvaredo:2019lmj}, where the physical momenta $p_i=\f{2\pi k_i}{L}$ are used.}
In the scaling limit $L\to+\infty$, $\ell\to+\infty$ with fixed ratio $x=\f{\ell}{L}$, as well as the limit that all the momenta $k_i$'s are large with fixed finite physical momenta $p_i=\f{2\pi k_i}{L}$ as $L\to+\infty$,\footnote{We thank Olalla Castro-Alvaredo, Cecilia De Fazio, Benjamin Doyon and Istv\'an Sz\'ecs\'enyi for explaining to us the precise limit they have worked in.} the R\'enyi entanglement entropy power is universal \cite{Castro-Alvaredo:2018dja,Castro-Alvaredo:2018bij,Castro-Alvaredo:2019irt,Castro-Alvaredo:2019lmj}
\be \label{CDDSuniversal}
\cF_{A,p_1^{r_1}p_2^{r_2}\cdots p_s^{r_s}}^{(n),\univ} = \prod_{i=1}^s \Big\{ \sum_{a=0}^{r_i} [ C_{r_i}^a x^a (1-x)^{{r_i}-a} ]^n \Big\},
\ee
where $C_{r_i}^a$ is the binomial coefficient.
{Note that the universal R\'enyi entanglement entropy power is only valid for finite $p_i$ in the limit $L\to+\infty$.}
We write the universal R\'enyi entanglement entropy power as $\cF_{A,p_1^{r_1}p_2^{r_2}\cdots p_s^{r_s}}^{(n),\univ}$ instead of $\cF_{A,k_1^{r_1}k_2^{r_2}\cdots k_s^{r_s}}^{(n),\univ}$ to remind the reader of the different limit that was used in \cite{Castro-Alvaredo:2018dja,Castro-Alvaredo:2018bij,Castro-Alvaredo:2019irt}.
The above formula is valid as long as either the correlation length of the model $\f1\D$ or the maximal de Broglie wavelength of the excited quasiparticles is much smaller than the sizes of the subsystems \cite{Castro-Alvaredo:2018dja,Castro-Alvaredo:2018bij,Castro-Alvaredo:2019irt}
\be \label{condition}
\min\Big[ \f1\D, \max_i\Big(\f{L}{|k_i|}\Big) \Big] \ll \min( \ell, L-\ell ),
\ee
but one does not need to impose both.
The universal R\'enyi entanglement entropy is independent of the model, the quasiparticle momenta and the connectedness of the subsystem.
In this letter we will report that there are nontrivial additional contributions to the universal R\'enyi entanglement entropy of quasiparticle excitations in the fermionic, bosonic, and XY chains when the momenta $k_i$ are general, i.e. that they can be small and/or close to each other.
%\red{We make the analytical calculations in discrete models, and this is why we find the additional terms that cannot be analytically deduced in continuous quantum field theories in \cite{Castro-Alvaredo:2018dja,Castro-Alvaredo:2018bij,Castro-Alvaredo:2019irt}.}

{\it{Fermionic chain}}:
We consider the chain of spinless fermions
\begin{multline}\label{fermionchain}
H = \sum_{j=1}^L \Big[ \lam \Big( a_j^\dag a_j - \f12 \Big) - \f12 ( a_j^\dag a_{j+1} + a_{j+1}^\dag a_j ) \\ - \f{\g}{2} ( a_j^\dag a_{j+1}^\dag + a_{j+1} a_j ) \Big],
\end{multline}
where the number of the total sites $L$ is an even integer.
In this letter we only consider the excitations in the Neveu-Schwarz sector, i.e.\ that we impose the anti-periodic boundary conditions on the fermions.
The Hamiltonian can be diagonalized in terms of the lowering and raising operators $c_k$, $c_k^\dag$ \cite{Lieb:1961fr,katsura1962statistical,pfeuty1970one}
\be
H = \sum_k \ve_k \Big( c_k^\dag c_k - \f12 \Big),
\ee
with the spectrum $\ve_k = \sr{ (\lam - \cos\f{2\pi k}{L})^2 + \g^2 \sin^2\f{2\pi k}{L} }$ and momenta $k=\f{1-L}{2},\cdots,-\f12,\f12,\cdots,\f{L-1}{2}$.
The ground state $|G\rag$ is annihilated by all the lowering operators $c_k|G\rag=0$,
and the excited states are generated by applying the raising operators with different momenta on the ground state
\be \label{XYk1k2cdotsksrag}
|k_1k_2\cdots k_s\rag = c_{k_1}^\dag c_{k_2}^\dag \cdots c_{k_s}^\dag |G\rag.
\ee

We first consider the single-interval case as shown in the left panel of FIG.~\ref{subsystems}.
In the extremely gapped limit $\lam\to+\infty$, the ground state has zero R\'enyi entanglement entropy $S_{A,G}^{(n)} = 0$.
In the limit $L\to+\infty$, $\ell\to+\infty$, $k_i\to+\infty$ with fixed $x=\f\ell{L}$, $p_i=\f{2\pi k_i}{L}$, the R\'enyi entanglement entropy power in the excited state $|k_1k_2\cdots k_s\rag$ takes a universal form \cite{Castro-Alvaredo:2018dja,Castro-Alvaredo:2018bij,Castro-Alvaredo:2019irt,Castro-Alvaredo:2019lmj}
\be \label{cFAk1k2cdotsksnuniv}
\cF_{A,p_1p_2\cdots p_s}^{(n),\univ} = [ x^n + (1-x)^n ]^s.
\ee
{In the extremely gapped limit $\lam\to+\infty$, the excited state $|k_1k_2\cdots k_s\rag$ can be written in terms of subsystem excitations. In the basis of the subsystem modes, the RDM is of finite dimension and we get analytically the R\'enyi entanglement entropy.
We call it the subsystem mode method, which is efficient for both analytical and numerical calculations, and we show details of the method in \cite{Zhang:2020dtd,WorkZR}.
The von Neumann and R\'enyi entanglement entropies of the ground state and excited states can also be calculated numerically from exact diagonalization, i.e. the correlation matrix method \cite{chung2001density}.}

\begin{figure}[ht]
  \centering
  \includegraphics[height=0.14\textwidth]{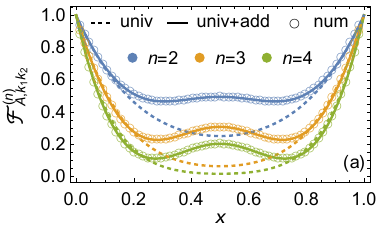}
  \includegraphics[height=0.14\textwidth]{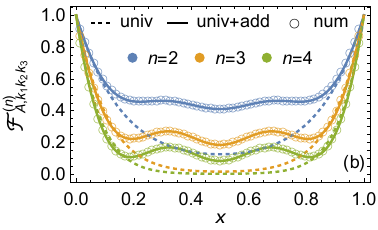}\\
  \includegraphics[height=0.14\textwidth]{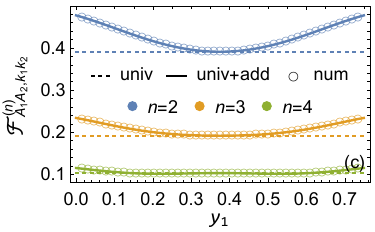}
  \includegraphics[height=0.14\textwidth]{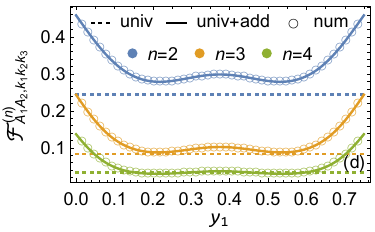}
  \caption{The single-interval (top) and double-interval (bottom) R\'enyi entanglement entropy powers in the double-particle (left) and triple-particle (right) states in the extremely gapped fermionic chain, including the universal results (dotted lines, ``univ''), the analytical results with additional terms (solid lines, ``univ+add''), and the numerical results (empty circles, ``num'').
  We use different colors for different R\'enyi indices $n$.
  We have used $\lam=+\infty$, $L=64$, $k_1=\f12$, $k_2=\f32$, $k_3=\f52$, $x_1=x_2=\f18$.}\label{fermionAA1A2}
\end{figure}

For the state with a single quasiparticle $|k\rag$, we get the same analytical and numerical results as the universal R\'enyi entanglement entropy. However,
for the state with two different quasiparticles $|k_1k_2\rag$ excited, we get the new result with additional contribution
\be \label{fermionZRk1k2}
\cF_{A,k_1k_2}^{(2)} = \cF_{A,p_1p_2}^{(2),\univ} + 8 x ( 1 - x) \a_{k_1-k_2}^2 + 4 \a_{k_1-k_2}^4,
\ee
where
\be\label{alphak}
\a_k = \f{\sin({\pi k\ell}/{L})}{L\sin({\pi k}/{L})}.
\ee
Similar formula can be also derived for $\cF_{A,k_1k_2}^{(n)}$ with a general integer $n$ as we show the details in \cite{Zhang:2020dtd}.
We have not taken into account any approximation except considering the extremely gapped limit, and so these analytical results are exact for any $L$, $\ell$, $k_1$, $k_2$. This is different from the limit in which the universal R\'enyi entanglement entropy power (\ref{cFAk1k2cdotsksnuniv}) was derived in \cite{Castro-Alvaredo:2018dja,Castro-Alvaredo:2018bij,Castro-Alvaredo:2019irt}, where all $L$, $\ell$, $k_1$, $k_2$, $|k_1-k_2|$ are large.
For a finite fixed $k=k_1-k_2$, the factor $\a_k$ is scale invariant in the limit $L\to+\infty$, $\ell\to+\infty$ with fixed $x=\f\ell{L}$
\be
\a_k = \f{\sin(\pi k x)}{\pi k}.
\ee
On the contrary, it is vanishing in the limit $|k|\to+\infty$.
For a large momentum difference $|k_1-k_2|$, the R\'enyi entanglement entropy power $\cF_{A,k_1k_2}^{(2)}$ (\ref{fermionZRk1k2}) approaches the universal R\'enyi entanglement entropy power $\cF_{A,p_1p_2}^{(2),\univ}$.
We compare the universal results of the R\'enyi entanglement entropy power, the results with nontrivial additional terms, and the numerical results in FIG.~\ref{fermionAA1A2}\red{(a)}.
There are non-negligible additional terms when the momentum difference $|k_1-k_2|$ is small.
The results with nontrivial additional terms match perfectly the numerical results.
We also calculated analytically the R\'enyi entanglement entropy power $\cF_{A,k_1k_2k_3}^{(n)}$ for the states with three different quasiparticles $|k_1k_2k_3\rag$. The results are reported in \cite{Zhang:2020dtd}.
In FIG.~\ref{fermionAA1A2}\red{(b)} we compare the universal results, the results with additional terms, and the numerical results.
The results with additional terms match perfectly the numerical ones.

We then consider the double-interval case, as shown in the right panel of FIG.~\ref{subsystems}. In the extremely gapped limit, we calculated the analytical R\'enyi entanglement entropy power in the single-particle, double-particle, and triple-particle states.
The double-interval R\'enyi entanglement entropy power in the single-particle state is the same as the universal result, while there are nontrivial additional contributions to the R\'enyi entanglement entropy power in the double-particle and triple-particle states.
We compare the universal results, the ones with additional terms, and the numerical results in FIG.~\ref{fermionAA1A2}\red{(c)} and \ref{fermionAA1A2}\red{(d)}.
There are perfect matches between the new results and the numerical ones.
Especially, note that the universal double-interval R\'enyi entanglement entropy power is independent of $y_1=\f{d_1}{L}$, however we see nontrivial dependence on $y_1$ for the new and the numerical results.

\begin{figure}[ht]
  \centering
  \includegraphics[height=0.14\textwidth]{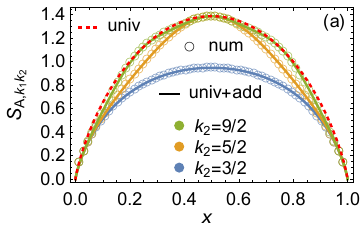}
  \includegraphics[height=0.14\textwidth]{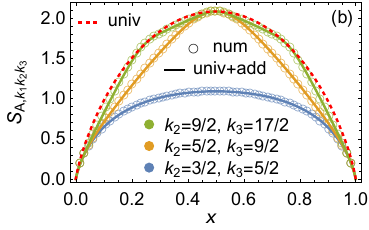}\\
  \caption{The universal single-interval von Neumann entanglement entropy (dotted lines), the new analytical results with additional corrections (solid lines) and the numerical results (empty circles) in the double-particle (left) and triple-particle (right) states in the extremely gapped fermionic chain.
  We use different colors for different momenta.
  We have used $\lam=+\infty$, $L=64$, $k_1=\f12$.}\label{XYEE}
\end{figure}

The universal R\'enyi entanglement entropy power (\ref{cFAk1k2cdotsksnuniv}) leads to the universal von Neumann entanglement entropy \cite{Castro-Alvaredo:2018dja,Castro-Alvaredo:2018bij,Castro-Alvaredo:2019irt,Castro-Alvaredo:2019lmj}
\be
S_{A,p_1p_2\cdots p_s}^\univ = s [ - x\log x -(1-x)\log(1-x) ].
\ee
We could analytically evaluate the excited state von Neumann entanglement entropy.
For example, in the double-particle state we get the single-interval von Neumann entanglement entropy
\bea
&& S_{A,k_1k_2} = - (x + \a_{k_1-k_2}) \log (x + \a_{k_1-k_2}) \nn\\
&& \phantom{S_{A,k_1k_2} =}
               - (x - \a_{k_1-k_2}) \log (x - \a_{k_1-k_2}) \\
&& \phantom{S_{A,k_1k_2} =}
               - (1 - x - \a_{k_1-k_2}) \log (1 - x - \a_{k_1-k_2}) \nn\\
&& \phantom{S_{A,k_1k_2} =}
               - (1 - x + \a_{k_1-k_2}) \log (1 - x + \a_{k_1-k_2}), \nn
\eea
with the definition of $\a_k$ (\ref{alphak}).
We compare the universal single-interval von Neumann entanglement entropy, the new results and the numerical results in the double-particle and triple-particle states, as shown in FIG.~\ref{XYEE}.
One could see the necessity of the additional contributions to the von Neumann entanglement entropy of the excited states with small momentum differences.

{\it{Bosonic chain}}:
Here we consider the periodic bosonic chain, i.e. the harmonic chain
\be
H = \f{1}{2} \sum_{j=1}^L \big[ p_j^2 + m^2 q_j^2 + (q_j-q_{j+1})^2 \big],
\ee
with the total number of sites $L$ being an even integer and the identification $q_{L+1}=q_1$.
The Hamiltonian in terms of the bosonic lowering and raising operators is
\be
H = \sum_k \ve_k \Big( b_k^\dag b_k +\f12 \Big),
\ee
with the spectrum $\ve_k = \sr{m^2+4\sin^2\f{\pi k}{L}}$ and momenta $k=1-\f{L}{2},\cdots,-1,0,1,\cdots,\f{L}{2}$.
The ground state $|G\rag$ is annihilated by all the lowering operators, i.e. $b_k | G \rag = 0$.
The excited states are constructed by applying various raising operators on the ground state
\be \label{k1r1k2r2cdotsksrsrag}
|k_1^{r_1}k_2^{r_2}\cdots k_s^{r_s}\rag = \f{( b^\dag_{k_1} )^{r_1} ( b^\dag_{k_2} )^{r_2} \cdots ( b^\dag_{k_s} )^{r_s}}{\sr{r_1!r_2!\cdots r_s!}}  | G \rag,
\ee
with general momenta $k_i$.
In the infinitely massive limit $m\to+\infty$, the ground state has vanishing R\'enyi entanglement entropy $S_{A,G}^{(n)} = 0$.
The universal excited R\'enyi entanglement entropy with large $k_i$ is (\ref{CDDSuniversal}).
{Similar to the fermionic chain, in the extremely gapped limit $m\to+\infty$, we write the excited state $|k_1^{r_1}k_2^{r_2}\cdots k_s^{r_s}\rag$ with general $k_i$ in terms of subsystem excitations, and calculate the excited state R\'enyi entanglement entropy power analytically from the subsystem mode method.
The excited state R\'enyi entanglement entropy can be also calculated from the wave function method \cite{Castro-Alvaredo:2018dja,Castro-Alvaredo:2018bij}.}

For both the single-interval and double-interval cases in the excited state of $r$ quasiparticles with equal momenta $|k^r\rag$, we get the same analytical and numerical results as the universal R\'enyi entanglement entropy power. %$\cF_{A,k^r}^{(n)} = \cF_{A,p^r}^{(n),\univ}$.
For a single interval in the state with two different quasiparticles $|k_1k_2\rag$, we get the R\'enyi entanglement entropy power with additional terms
\be
\cF_{A,k_1k_2}^{(2)} = \cF_{A,p_1p_2}^{(2),\univ}
      + 4 (1 - 2 x)^2 \a_{k_1-k_2}^2
      + 4 \a_{k_1-k_2}^4,
\ee
where $\a_{k}$ is defined in (\ref{alphak}).
Similar formulas can be also derived for $\cF_{A,k_1k_2}^{(n)}$ with larger $n$ and also for a state with three different quasiparticles $|k_1k_2k_3\rag$, we calculated $\cF_{A,k_1k_2k_3}^{(n)}$ with nontrivial additional contributions to the universal results in all the cases.
The double-interval R\'enyi entanglement entropy power in the double-particle and triple-particle states $\cF_{A_1A_2,k_1k_2}^{(n)}$, $\cF_{A_1A_2,k_1k_2k_3}^{(n)}$ were also calculated analytically which in all the cases we found additional contributions to the universal term.
Although not shown here the results of the R\'enyi entanglement entropy power with nontrivial additional terms match perfectly with the results from the parafunction method \cite{Zhang:2020dtd}.

{\it{XY chain}}:
The XY chain has the Hamiltonian
\be \label{XYchain}
H = - \sum_{j=1}^L \Big( \f{1+\g}{4} \s_j^x\s_{j+1}^x + \f{1-\g}{4} \s_j^y\s_{j+1}^y + \f\lam2 \s_j^z \Big),
\ee
with the Pauli matrices $\s_j^{x,y,z}$.
It is related to the fermionic chain (\ref{fermionchain}) by the Jordan-Wigner transformation \cite{Lieb:1961fr,katsura1962statistical,pfeuty1970one}
\be
a_j = \Big(\prod_{i=1}^{j-1}\s_i^z\Big) \s_j^+, ~~
a_j^\dag = \Big(\prod_{i=1}^{j-1}\s_i^z\Big) \s_j^-,
\ee
with $\s_j^\pm \equiv \f12 ( \s_j^x \pm \ii \s_j^y )$ and properly taking care of the boundary conditions.
The ground and excited states in the XY chain can be constructed in the same way as those in the fermionic chain.

\begin{figure}[ht]
  \centering
  \includegraphics[height=0.2\textwidth]{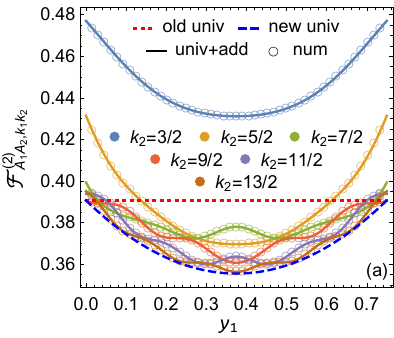}
  \includegraphics[height=0.2\textwidth]{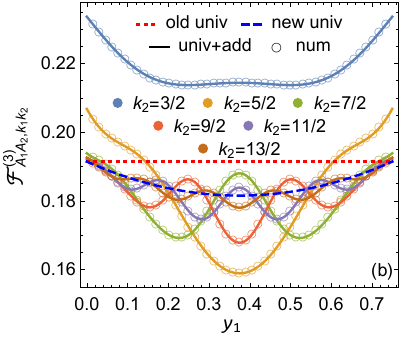}
  \caption{The double-interval R\'enyi entanglement entropy power in the double-particle state in the extremely gapped XY chain, including the old universal results (red dotted lines), the new universal results (blue dashed lines), the new universal results plus additional terms (solid lines), and the numerical results (empty circles).
  We also use different colors for different momenta.
  We have set $\lam=+\infty$, $L=64$, $k_1=\f12$, $x_1=x_2=\f18$.}\label{XYA1A2}
\end{figure}

For a single interval, the R\'enyi entanglement entropy power is the same as that of the fermionic chain.
However, it is different for the double-interval case \cite{Alba:2009ek,Igloi:2009On}.
In the extremely gapped limit, we use the subsystem mode method and obtain exactly the double-interval R\'enyi entanglement entropy power $\cF_{A_1A_2,k_1k_2}^{(n)}=\cF_{A_1A_2,p_1p_2}^{(n),\univ}+\d\cF_{A_1A_2,k_1k_2}^{(n)}$.
Especially, we get for example the new universal results
\bea \label{ZRuniversal1}
&& \cF_{A_1A_2,p_1p_2}^{(2),\univ} = (x^2 + y^2)^2 - 16 x_1 x_2 y_1 y_2, \\
\label{ZRuniversal2}
&& \cF_{A_1A_2,p_1p_2}^{(3),\univ} = (x^3 + y^3)^2 - 24 x_1 x_2y_1 y_2 x y,
\eea
which are different from (\ref{CDDSuniversal}).
The double-interval R\'enyi entanglement entropy was calculated numerically in \cite{Furukawa:2008uk,Facchi:2008Entanglement,Alba:2009ek,Igloi:2009On,Fagotti:2010yr}, and we adopt the effective method in \cite{Fagotti:2010yr}.
The analytical and numerical results of the R\'enyi entanglement entropy power are compared in FIG.~\ref{XYA1A2}, and we see perfect matches.
{In FIG.~\ref{XYA1A2} we just choose $x_1=x_2=\f18$ as an example, and there are similar results for other values of $x_1$, $x_2$.}
In the limit of large momentum difference $|k_1-k_2|$, the numerical results approach to the new universal results (\ref{ZRuniversal1}) and (\ref{ZRuniversal2}) instead of the old one (\ref{CDDSuniversal}).
The universal result (\ref{CDDSuniversal}) has an elegant quantum information interpretation in terms of entangled qubits \cite{Castro-Alvaredo:2018dja,Castro-Alvaredo:2018bij,Castro-Alvaredo:2019irt,Castro-Alvaredo:2019lmj}, which breaks down for the double-interval R\'enyi entanglement entropy in the double-particle state in the XY chain (\ref{ZRuniversal1}) and (\ref{ZRuniversal2}) even in the limit that all the momenta and the momentum difference are large.
Although quite remarkable, this is not surprising as the local excitations in the XY chain are the same as the ones in neither the fermionic nor the bosonic chain.
In fact, as stated in \cite{Castro-Alvaredo:2018dja,Castro-Alvaredo:2018bij}, the validity of the universal R\'enyi entanglement entropy therein requires that the quasiparticles are localized quantum excitations, while it is not the case in the XY chain.

{\it{Discussion}}:
In this letter we have used the extremely gapped fermionic and bosonic Hamiltonians and calculated exactly the R\'enyi entanglement entropy in the quasiparticle excited states.
In addition to the universal terms  derived for large $k_i$ in \cite{Castro-Alvaredo:2018dja,Castro-Alvaredo:2018bij,Castro-Alvaredo:2019irt,Castro-Alvaredo:2019lmj} we found extra terms that cannot be ignored in the limit of small difference between the excited momenta.
Technically, the additional term comes from the (anti)commuting relations of the local modes, and we could call them ``exchange term''. The exchange term is larger when the quasiparticles have closer but distinct momenta.
These extra terms are also valid beyond the regime they are derived.
For example, we realized that for the free fermionic Hamiltonian (\ref{fermionchain}) and the XY chain (\ref{XYchain}) the result (\ref{fermionZRk1k2}) remains intact also for the Hamiltonian with general parameters $\g,\lam$ and momenta $k_1,k_2$ as long as the condition (\ref{condition}) is satisfied.
It is expected that if one considers a generic Hamiltonian $H=H_g+\lambda\sum_{j=1}^L(a_j^{\dagger}a_j-\frac{1}{2})$ in the large $\lambda$ limit one recovers the result (\ref{fermionZRk1k2}).
We also note that the magnon states considered in the supplemental material of \cite{Castro-Alvaredo:2018dja} leads to the same result as ours after considering the limit that $k_i$'s are general.
The same argument can be also applied for bosonic systems which supports the idea of the universality of our results.
Our universality argument can be also generalized to the double interval situation in the case of the fermionic and bosonic chains.
However, the double interval case in the XY chain ends up to have a completely novel behavior. Instead of the universal results discovered in \cite{Castro-Alvaredo:2018dja} we found a new universal term with new additional terms.
The difference from the fermionic chain is due to the fact that the local degrees of freedom of the XY chain are the Pauli matrices not the spinless fermions.
We expect the equations (\ref{ZRuniversal1}) and (\ref{ZRuniversal2}) to be also universal as far as one considers double interval in quantum spin-$\f12$ chains.

{Although we have derived the analytical R\'enyi entanglement entropy in the extremely gapped limit, we also checked numerically that the R\'enyi entanglement entropy powers with nontrivial additional terms in the bosonic and the fermionic chains also apply to the slightly gapped and even critical chains as long as all the excited quasiparticles have large momenta.
As $L\to+\infty$, in the fermionic and XY chains we fit numerically for several examples and get
\be
| 1 - {\cF_{A,K}^{(n)}({\rm finite}~\lambda)}/{\cF_{A,K}^{(n)}(\lambda=+\infty)} | \sim \f{1}{L},
\ee
and in the bosonic chain we get
\be
| 1 - {\cF_{A,K}^{(n)}({\rm finite}~m)}/{\cF_{A,K}^{(n)}(m=+\infty)} | \sim \f{1}{L^2}.
\ee
We show more details in \cite{Zhang:2020dtd}.}%

One highlight of the paper is the momentum dependence of the R\'enyi and entanglement entropies in the quasiparticle excited states, improving the results in \cite{Pizorn2012Universality,Berkovits2013Twoparticle,Molter2014Bound,Castro-Alvaredo:2018dja,Castro-Alvaredo:2018bij,Castro-Alvaredo:2019irt}.
The momentum independence of the von Neumann entanglement entropy has inspired the studies of local quantum quenches in \cite{Eisler:2019gnr,Gruber:2020nzw}, which can possibly be revisited with the new results in this letter.
%In fact, in a special but still physical circumstance, all the quantum quenches would be trivial if the quasiparticles are independent, as we show in the supplemental material of this letter.
The quasiparticle excited states can be engineered experimentally \cite{Jurcevic:2014qfa,Jurcevic:2015inc}, and it is interesting to measure the quasiparticle excite state R\'enyi entanglement entropy and compare the data with our momentum-dependent results.
The universal R\'enyi and entanglement entropies we have obtained can also be used as a benchmark to other methods, for example the tensor network techniques \cite{Haegeman:2011lcd,Moudgalya:2018kqm,Moudgalya:2018nuz,Tu:2021tjn}.
The techniques to calculate the R\'enyi and entanglement entropies are also useful to calculate the subsystem Schatten and trace distances \cite{WorkZR}.

{\it{Acknowledgements}}:
We thank Pasquale Calabrese for careful reading of a previous version of the draft, encouragement, and valuable discussions, comments and suggestions.
We also thank D\'avid Horv\'ath, Gianluca Lagnese and Sara Murciano for helpful discussions.
We are indebted to Olalla Castro-Alvaredo, Cecilia De Fazio, Benjamin Doyon and Istv\'an Sz\'ecs\'enyi for important comments and clarifications.
MAR thanks CNPq and FAPERJ (grant number 210.354/2018) for partial support.
JZ acknowledges support from ERC under Consolidator grant number 771536 (NEMO).

\providecommand{\href}[2]{#2}\begingroup\raggedright\endgroup

%\bibliographystyle{D:/00.bibx/apsrev4-1}
%\bibliographystyle{D:/00.bibx/JHEP}
%\bibliographystyle{D:/00.bibx/JHEPx}
%\bibliographystyle{D:/00.bibx/JHEPy} %% for PRL with arxiv numbers but without tiltes
%\bibliographystyle{D:/00.bibx/JHEPz} %% for PRL with arxiv numbers but without tiltes or "arXiv:"
%\bibliographystyle{D:/00.bibx/eplbib} %% for EPL
%\bibliography{D:/00.bibx/2021,D:/00.bibx/2020,D:/00.bibx/2019,D:/00.bibx/2018,D:/00.bibx/1960,D:/00.bibx/1970,D:/00.bibx/1980,D:/00.bibx/1990,D:/00.bibx/1995,D:/00.bibx/1996,D:/00.bibx/1997,D:/00.bibx/1998,D:/00.bibx/1999,D:/00.bibx/2000,D:/00.bibx/2001,D:/00.bibx/2002,D:/00.bibx/2003,D:/00.bibx/2004,D:/00.bibx/2005,D:/00.bibx/2006,D:/00.bibx/2007,D:/00.bibx/2008,D:/00.bibx/2009,D:/00.bibx/2010,D:/00.bibx/2011,D:/00.bibx/2012,D:/00.bibx/2013,D:/00.bibx/2014,D:/00.bibx/2015,D:/00.bibx/2016,D:/00.bibx/2017,D:/00.bibx/book,D:/00.bibx/work,D:/00.bibx/thesis}

\end{document}